\documentclass{ifacconf}

\usepackage{graphicx}      
\usepackage{subcaption}
\usepackage{amsmath}

\usepackage{dsfont}
\usepackage{xcolor}
\usepackage{upgreek}
\usepackage{enumerate}
\usepackage{amsfonts}
\usepackage{amssymb}
\usepackage{textcomp}
\usepackage{natbib}        
\usepackage{mdframed}
\usepackage{tikz}
\usepackage{mathabx}
\usepackage{tkz-berge}
 \usepackage{algpseudocode}
\usetikzlibrary{calc}
\usetikzlibrary{shapes,fit}
\usepackage{verbatim}
\usepackage{setspace}
\usetikzlibrary{shapes,arrows}

\newtheorem{theorem}{Theorem}[section]

\newtheorem{lemma}[theorem]{Lemma}
\newtheorem{assumption}[theorem]{Assumption}
\newtheorem{definition}[theorem]{Definition}

\newtheorem{remark}[theorem]{Remark}

\DeclareMathOperator{\diag}{diag}

\begin{document}
\begin{frontmatter}

\title{Formation Control for Multi-Agent Systems with Connectivity Preservation and Event-Triggered Controllers \thanksref{footnoteinfo}}

\thanks[footnoteinfo]{This work was supported by the
Knut and Alice Wallenberg Foundation, the  Swedish Foundation for Strategic Research, and the Swedish Research Council.}

\author{Xinlei Yi, Jieqiang Wei, Dimos V. Dimarogonas, Karl H. Johansson}

\address{The ACCESS Linnaeus Centre and Electrical Engineering, \\KTH Royal Institute of Technology, 100 44, Stockholm, Sweden \\(e-mail: xinleiy, jieqiang, dimos, kallej@kth.se).}

\begin{abstract}\label{s:Abstract}
In this paper, event-triggered controllers and corresponding algorithms are proposed to establish the formation with connectivity preservation for multi-agent systems. Each agent needs to update its control input and to broadcast this control input together with the relative state information to its neighbors at its own triggering times, and to receive information at its neighbors' triggering times. Two types of system dynamics, single integrators and double integrators, are considered. As a result, all agents converge to the formation exponentially with connectivity preservation, and Zeno behavior can be excluded. Numerical simulations show
the effectiveness of the theoretical results.
\end{abstract}


\end{frontmatter}

\section{Introduction}
In the last two decades, numerous contributions have been given in the study of distributed cooperative control, including consensus control and formation control, etc, for multi-agent systems, see e.g., \cite{Jadbabaie2003}, \cite{Olfati2004}, \cite{Moreau2004}, \cite{Ren2005}, \cite{Liu2011}. In these contributions, there are two common assumptions that are assumed expressly or implicitly. One is that all agents must maintain a (strongly or weakly) connected communication graph either for all time or in an average sense. Another one is that the information can be continuously transmitted between agents or each agent can continuously sense the relative information, such as relative positions. However, it is not straightforward to verify the connectivity since the states of agents change during the evolution of the system and thus the connection between two agents may changes. Moreover, it may be also impractical to require continuous sensing or broadcasting  in real applications.

Connectivity preservation is a fundamental problem in the study of many areas, such as mobile robots.
Normally, mobile robots have limited communication, including sensing and broadcasting, capabilities and one robot can only communicate with other robots that are within close enough distance. It is natural to model the interaction between a group of mobile robots with limited communication radii as a multi-agent system and the corresponding communication graph is determined by their relative distances. 
Motivated by this, many researchers studied distributed cooperative control for multi-agent systems with connectivity preservation, see e.g., \cite{Spanos2004}, \cite{Dimarogonas2007}, \cite{Ji2007}, \cite{Zavlanos2007}, \cite{Ajorlou2010}, \cite{Kan2012}, \cite{Boskos2015}. Different from this paper, in these papers continuous sensing or broadcasting is needed.

One the other hand, one common method to avoid continuous broadcasting and sensing is using periodical sampling. However the drawbacks of this method are that it normally requires a large amount of agents take action in a synchronous manner and it is not efficient. Motivated by the future trend that agents can be equipped with embedded microprocessors with limited resources to transmit and collect information, event-triggered control was introduced by \cite{aastrom1999comparison}, \cite{tabuada2007event}. Instead of using the continuous state to realize desired convergence properties, the
control in the event-triggered control strategy is piecewise constant between
the triggering times which need to be determined. A key challenge in event-triggered control is how to design the  triggering conditions
to determine the triggering times, and to exclude Zeno behavior. For continuous-time multi-agent systems, the Zeno behavior is the behavior that there are infinite number of triggers in a
finite time interval, see \cite{johansson1999regularization}. To name a few existing works on multi-agent systems with event-triggered controllers, see e.g., \cite{dimarogonas2012distributed}, \cite{seyboth2013event}, \cite{meng2015periodic}, \cite{nowzari2016distributed}. Different from this paper, these papers do not consider the  connectivity preservation.

In this paper, we study formation control for multi-agent systems with connectivity preservation and event-triggered controller. We propose distributed conditions for agents to determine their triggering times and two corresponding algorithms for each agent to avoid continuous checking of its own triggering condition. Moreover, it is only at its triggering times that each agent needs to update its control input by sensing the relative positions between itself and its
neighbors, to broadcast its triggering information, including current triggering time, the relative states and control input at this time, to its neighbors. In other words, all agents only need to do broadcasting and sensing at their triggering times and this occurs  not in synchronous manner. As a result, all agents converge to a formation exponentially with connectivity preservation. In addition, Zeno behavior can be excluded by proving that the inter-event times are lower bounded by a positive constant. Two related existing papers are \cite{yu2012formation}, \cite{fan2015connectivity}. However, \cite{yu2012formation} didn't explicitly exclude the Zeno behavior, but it is well known that such behavior can cause problem, see \cite{johansson1999regularization}. And it was under the assumption that no agent exhibits Zeno behavior, \cite{fan2015connectivity} proved asymptotic  rendezvous can be achieved.

The rest of this paper is organized as follows. Section \ref{secpreliminaries} introduces the preliminaries. The main results are stated in Section \ref{secsingle} and Section \ref{secdouble}. Simulations are given in Section \ref{secsimulations}. Finally, the paper is concluded in Section \ref{secconclusions}.

\noindent {\bf Notions}: $\|\cdot\|$ represents the Euclidean norm for
vectors or the induced 2-norm for matrices. $|S|$ is the cardinality of set $S$.  ${\bf 1}_n$ denotes the column
vector with each component 1 and dimension $n$. $I_n$ is the $n$ dimension identity matrix. $\rho(\cdot)$ stands for
the spectral radius for matrices and $\rho_2(\cdot)$ indicates the minimum
positive eigenvalue for matrices having positive eigenvalues. Given two
symmetric matrices $M,N$, $M\ge N$ means $M-N$ is a positive semi-definite matrix. The notation $A\otimes B$ means the Kronecker product of matrices $A$ and $B$.
\section{PRELIMINARIES}\label{secpreliminaries}

In this section we review some definitions and results from algebraic graph theory and formation control, for more details see \cite{mesbahi2010graph}.

\subsection{Algebraic Graph Theory}
For a {\it undirected graph} $\mathcal G=(\mathcal V,\mathcal E, A)$ with $n$ nodes (vertices), the set of
nodes $\mathcal V =\{1,\dots,n\}$, set of edges $\mathcal E
\subseteq \mathcal V \times \mathcal V$, and the {\it adjacency matrix}
$A =(a_{ij})\in\mathbb{R}^{n\times n}$ given by $a_{ij}=a_{ji}=1$ if $(i, j)\in \mathcal E$, and  $a_{ij}=0$ otherwise. Let $N_i=\{j\in\mathcal V|(i,j)\in\mathcal E\}$ and $\deg(i)=\sum\limits_{j=1}^{n}a_{ij}$ denotes the {\it neighbors} and {\it degree} of agent $i$, respectively. The {\it degree matrix} of graph $\mathcal G$ is defined as $Deg=\diag([\deg(1), \cdots, \deg(n)])$. The {\it Laplacian matrix} is defined as $L=Deg-A$.

Similarly, we can define the {\it weighted graph} $\mathcal G=(\mathcal V,\mathcal E, \Omega)$ with $\Omega =(\omega_{ij}) \in\mathbb{R}^{n\times n}$ is a nonnegative symmetric matrix and $\omega_{ij}>0$ if $(i, j)\in \mathcal E$. And we denote the corresponding {\it Laplacian matrix} as $L_\omega$. 

A {\it path} of length $k$ between agent $i$ and agent $j$ is a subgraph with distinct agents $i_0=i,\dots,i_k=j\in\mathcal V$ and edges $(i_j,i_{j+1}),~j=0,\dots,k-1$. A graph is {\it connected} if there is a path between any two agents.

Let $m$ denotes the number of edges in  $\mathcal G$, i.e., $m=|\mathcal E(\mathcal G)|$ and label the edges in $\mathcal G$ as $e_1,\cdots,e_m$. And denote $W=\diag\{\omega(e_1),\cdots,\omega(e_m)\}$, where $\omega(e_k)=\omega_{ij}$ with $e_k$ being the label of edge $(i,j)$. After assigning a direction to each edge, the $n\times m$ {\it incidence matrix} $D=(d_{ij})$ is defined as
\begin{align}
d_{ij}=\begin{cases}
-1&\text{if agent } i \text{ is the tail of edge } e_j,\\
1&\text{if agent } i \text{ is the head of edge } e_j,\\
0&\text{otherwise}.
\end{cases}
\end{align}

\begin{lemma}
$DD^{\top}$ is independent of the labels and orientations given to $\mathcal G$, and $DD^{\top}=L, DWD^{\top}=L_\omega$.
\end{lemma}

\begin{lemma}\label{lemma2}
Let $K_n=I_n-\frac{1}{n}\bf 1_n\bf 1^{\top}_n$ and assume $\mathcal G$ is connected, then $K_nL_\omega=L_\omega$, $\rho_2(DD^{\top})>0$, $\rho(K_n)=1$ and
\begin{align}\label{DDK}
0\le\rho_2(DD^{\top})K_n\le DD^{\top}.
\end{align}
\end{lemma}
\begin{pf}
$K_nL_\omega=L_\omega$ is straightforward.
$\rho_2(DD^{\top})>0$ is also straightforward since  $\mathcal G$ is connected.

From Ger\v{s}gorin disc theorem, we know that $\rho(K_n)\le1$. From $det(K_n-I_n)=0$, we know that 1 is an eigenvalue of $K_n$. Thus $\rho(K_n)=1$.

$K_n\ge0$ is also straightforward since we can regard $K_n$ as the Laplacian matrix of a complete graph.

For any $y\in\mathbb{R}^n$, there exist $y_0\in\mathbb{R}^n$ and $\zeta\in\mathbb{R}$ such that $y=y_0+\zeta{\bf1}_n$ and $y_0\bot{\bf1}_n$.
From
\begin{align*}
\rho_2(DD^{\top})=\min_{x\bot {\bf1}_n,x\neq0}\frac{x^{\top}DD^{\top}x}{\|x\|^2},~
\rho(K_n)=\max_{x\neq0}\frac{x^{\top}K_nx}{\|x\|^2},
\end{align*}
we get
\begin{align*}
y^{\top}DD^{\top}y=y^{\top}_0DD^{\top}y_0\ge\rho_2(DD^{\top})\|y_0\|^2,
\end{align*}
and
\begin{align*}
y^{\top}K_ny=y^{\top}_0K_ny_0\le\rho(K_n)\|y_0\|^2=\|y_0\|^2.
\end{align*}
Thus
\begin{align*}
y^{\top}DD^{\top}y\ge\rho_2(DD^{\top})\|y_0\|^2\ge\rho_2(DD^{\top})y^{\top}K_ny.
\end{align*}
In conclusion, (\ref{DDK}) holds.
\end{pf}

\subsection{Formation Control Problem}
Given a connected graph $\mathcal G$, let $d_{ij}\in\mathbb{R}^{p}$ denotes the {\it desired internode displacement} of edge $(i,j)\in\mathcal E(\mathcal G)$. Denote $\Phi=\{(\tau_1^{\top},\dots,\tau_n^{\top})^{\top}\in\mathbb{R}^{np}|\tau_i-\tau_j=d_{ij},\text{ for all }(i,j)\in\mathcal E(\mathcal G)\}$. We call the set of desired internode displacements $\{d_{ij},(i,j)\in\mathcal E(\mathcal G)\}$ a formation associated with $\mathcal G$ and we say it is {\it feasible} if $\Phi\neq\emptyset$.

Consider a multi-agent system with $n$ agents corresponding to the $n$ nodes in this graph. 
Let $x_i(t)\in\mathbb{R}^{p}$ denotes the position of agent $i$ at time $t\ge0$.  A group of agents are said to converge to a {\it formation} if $$\lim_{t\to\infty}x_i(t)-x_j(t)=d_{ij}$$ for all $(i,j)\in\mathcal E(\mathcal G)$.

In practice, agents normally have limited communication capabilities. For simplicity we assume all agents have the same communication radius $\Delta>0$. We say the graph $\mathcal G$ and the multi-agent system are {\it consistent} if  $\|x_i(t)-x_j(t)\|\le\Delta$ for all $(i,j)\in\mathcal E(\mathcal G)$ and all time $t\ge0$. Namely, the communication channels are kept for all time. Notice here we assume the following.
\begin{assumption}\label{assum1}
$\{d_{ij},(i,j)\in\mathcal E(\mathcal G)\}$, a formation associated with $\mathcal G$, is feasible and $\|d_{ij}\|<\Delta$, $\forall (i,j)\in\mathcal E(\mathcal G)$.
\end{assumption}
\begin{definition}
A group of agents associated with a graph $\mathcal G$ are said to converge to a formation with connectivity preservation if they converge to a formation while the graph $\mathcal G$ remains consistent with their dynamics.
\end{definition}

\section{Formation Control with Connectivity Preservation: Single Integrators}\label{secsingle}
In this section, we consider the case when the dynamics of agents can be modeled as single integrators given by
\begin{align}\label{single}
\dot{x}_i(t)=u_i(t),~i\in\mathcal V.
\end{align}
From Assumption \ref{assum1}, we know $\Phi\neq\emptyset$. Choose $(\tau_1^{\top},\dots,\\\tau_n^{\top})^{\top}\in\Phi$ and let $y_i(t)=x_i(t)-\tau_i$ for $i=1,\dots,n$. Then, we can rewrite the above multi-agent system as
\begin{align}\label{singley}
\dot{y}_i(t)=u_i(t),~i\in\mathcal V.
\end{align}

For $\|y_i-y_j\|<\Delta-\|d_{ij}\|$, the edge-tension function $\nu_{ij}$ (introduced first in \cite{Ji2007}) is defined as
\begin{align}
\nu_{ij}(\Delta,y)=\begin{cases}
\frac{\|y_i-y_j\|^2}{\Delta-\|d_{ij}\|-\|y_i-y_j\|},&\text{if }(i,j)\in\mathcal E(\mathcal G),\\
0,&\text{otherwise.}
\end{cases}
\end{align}
with
\begin{align*}
\frac{\partial\nu_{ij}(\Delta,y)}{\partial y_i}=
\frac{2\Delta-2\|d_{ij}\|-\|y_i-y_j\|}{(\Delta-\|d_{ij}\|-\|y_i-y_j\|)^2}(y_i-y_j),
\end{align*}
if $(i,j)\in\mathcal E(\mathcal G)$, and $\frac{\partial\nu_{ij}(\Delta,y)}{\partial y_i}=0$ otherwise.

We denote as $\omega_{ij}(t)$ the weight coefficient of the partial derivative of $\nu_{ij}$ with respect to $y_i$ as above, i.e.,
\begin{align}
\omega_{ij}(t)=\begin{cases}
\frac{2\Delta-2\|d_{ij}\|-\|y_i(t)-y_j(t)\|}{(\Delta-\|d_{ij}\|-\|y_i(t)-y_j(t)\|)^2},&\text{if }(i,j)\in\mathcal E(\mathcal G),\\
0,&\text{otherwise.}
\end{cases}
\end{align}
Note that $\omega_{ij}(t)$ can also be written as a function of $x_i(t)$ and $x_j(t)$ since $y_i(t)-y_j(t)=x_i(t)-x_j(t)-d_{ij}$.

In order to reduce the overall need of sensing and system updates, we use the following event-triggered controller 
\begin{align}
&u_i(t)=\sum_{j\in N_i}-\omega_{ij}(t^i_{k_i(t)})(y_i(t^i_{k_i(t)})-y_j(t^i_{k_i(t)}))\label{uysingle}\\
=&\sum_{j\in N_i}-\omega_{ij}(t^i_{k_i(t)})(x_i(t^i_{k_i(t)})-x_j(t^i_{k_i(t)})-d_{ij}),\label{uxsingle}
\end{align}
where $k_{i}(t)=arg\max_{k}\{t^{i}_{k}\le t\}$ with the increasing time sequence $\{t_{k}^{i}\}_{k=1}^{\infty}$, which is named as {\em triggering times}. We assume $t_{1}^{i}=0$, for all $i\in\mathcal V$.
One can see that the control input (\ref{uxsingle}) only updates at the triggering times.

In the following theorem, we will give triggering conditions to determine the triggering times such that the formation with connectivity preservation can be established and the Zeno behavior can be excluded.
\begin{theorem}\label{thm1}
Given a graph $\mathcal G$ which is undirected and connected and a formation associated with $\mathcal G$ which satisfies Assumption \ref{assum1}.
Consider the multi-agent system (\ref{single}) with event-triggered controller (\ref{uxsingle}) associated with $\mathcal G$. Assume that at the initial time $t=0$,
\begin{align}\label{initial}
\|x_i(0)-x_j(0)-d_{ij}\|=\|y_i(0)-y_j(0)\|<\Delta-\|d_{ij}\|,
\end{align}
for all $(i,j)\in\mathcal E(\mathcal G)$.  Given $\alpha>0$, and $0<\beta<\beta_0$ with $\beta_0=\frac{\rho_2(DD^{\top})}{\Delta_0}$ and $\Delta_0=\max_{(i,j)\in\mathcal E(\mathcal G)}\Delta-\|d_{ij}\|$, and given the first triggering time $t^i_1=0$, agent $i$ determines the triggering times $\{t^i_k\}_{k=2}^{\infty}$ by
\begin{align}\label{triggersingle}
t^i_{k+1}=\max_{r\ge t^i_k}\{r:\|e_i(t)\|\le \alpha e^{-\beta t},\forall t\in[t^i_k,r]\}
\end{align}
where
\begin{align*}
e_i(t)=&\sum_{j\in N_i}\omega_{ij}(t)(x_i(t)-x_j(t)-d_{ij})\\
&-\sum_{j\in N_i}\omega_{ij}(t^i_{k_i(t)})(x_i(t^i_{k_i(t)})-x_j(t^i_{k_i(t)})-d_{ij}).
\end{align*}
Then the multi-agent system (\ref{single}) with event-triggered controller (\ref{uxsingle}) converges to the formation  exponentially with connectivity preservation, and there is no Zeno behavior.
\end{theorem}
\begin{pf}
This theorem holds if we can prove that
\begin{enumerate}
  \item $\|x_i(t)-x_j(t)\|\le\Delta, \forall (i,j)\in\mathcal E(\mathcal G), \forall t\ge0$;
  \item $\lim_{t\to\infty}x_i(t)-x_j(t)=d_{ij}, \forall(i,j)\in\mathcal E(\mathcal G)$,  exponentially;
  \item there exists $\xi_i>0$, such that $t^i_{k+1}-t^i_k\ge\xi_i$, $\forall i\in\mathcal V$ and $\forall k=1,2,\dots$.
\end{enumerate}

{\bf (i)} We define the total tension energy of $\mathcal G$ as
\begin{align}
\nu(\Delta,y)=\frac{1}{2}\sum_{i=1}^{n}\sum_{j\in N_i}\nu_{ij}(\Delta,y),
\end{align}
where $y(t)=(y_1^{\top}(t),\dots,y_n^{\top}(t))^{\top}$. The time derivative of $\nu(\Delta,y(t))$ along the trajectories of the multi-agent system (\ref{singley}) with event-triggered controller (\ref{uysingle}) is
\begin{align}
&\dot{\nu}(\Delta,y(t))=\sum_{i=1}^{n}\sum_{j\in N_i}\Big[\frac{\partial\nu_{ij}(\Delta,y)}{\partial y_i}\Big]^{\top}\Big|_{y=y(t)}\dot{y}_i(t)\nonumber\\
=&\sum_{i=1}^{n}\sum_{j\in N_i}[\omega_{ij}(t)(y_i(t)-y_j(t))]^{\top}\nonumber\\
&\times[e_i(t)-\sum_{j\in N_i}\omega_{ij}(t)(y_i(t)-y_j(t))]\nonumber\\
\le&-\|L_\omega y(t)\|^2+a\|L_\omega y(t)\|^2+\frac{1}{4a}\sum_{i=1}^{n}\|e_i(t)\|^2,
\end{align}
for all $0<a<1$. From (\ref{triggersingle}), we know that
\begin{align}
\|e_i(t)\|\le\alpha e^{-\beta t}, \forall t\ge0.
\end{align}
Hence
\begin{align*}
\dot{\nu}(\Delta,y(t))\le\frac{n\alpha^2}{4a}e^{-2\beta t}, \forall t\ge0.
\end{align*}
Thus
\begin{align}
\nu(\Delta,y(t))\le&\nu(\Delta,y(0))+\frac{n\alpha^2}{8a\beta}[1-e^{-2\beta t}]\nonumber\\
\le&k_{\nu}, \forall t\ge0,
\end{align}
where
\begin{align}
&k_{\nu}=\nu(\Delta,y(0))+\frac{n\alpha^2}{8a\beta}\nonumber\\
=&\frac{1}{2}\sum_{i=1}^{n}\sum_{j\in N_i}\frac{\|x_i(0)-x_j(0)-d_{ij}\|^2}{\Delta-\|d_{ij}\|-\|x_i(0)-x_j(0)-d_{ij}\|}+\frac{n\alpha^2}{8a\beta}.
\end{align}
Then, for any $(i,j)\in\mathcal E(\mathcal G)$ and $t\ge0$, we have
\begin{align*}
\nu_{ij}(\Delta,y(t))
\le2\nu(\Delta,y(t))\le 2k_{\nu}.
\end{align*}
Hence
\begin{align}\label{lengthys}
\|y_i(t)-y_j(t)\|\le k_{ij},
\end{align}
where
\begin{align}
k_{ij}=-k_\nu+\sqrt{k^2_\nu+2k_\nu(\Delta-\|d_{ij}\|)}<\Delta-\|d_{ij}\|.\label{kij}
\end{align}
Then, we have
\begin{align}
&\|x_i(t)-x_j(t)\|=\|x_i(t)-\tau_i-(x_j(t)-\tau_j)+d_{ij}\|\nonumber\\
\le&\|y_i(t)-y_j(t)\|+\|d_{ij}\|
\le k_{ij}+\|d_{ij}\|<\Delta,
\end{align}
and thus connectivity maintenance is established.

{\bf (ii)} Let $e(t)=(e_1^{\top}(t),\dots,e_n^{\top}(t))^{\top}$, $\bar{y}(t)=\frac{1}{n}\sum_{i=1}^{n}y_i(t)$ and $\delta(t)=y(t)-{\bf1}_n\otimes\bar{y}(t)=(K_n\otimes I_p)y(t)$. We consider the  Lyapunov candidate
\begin{align}
V(t)=\frac{1}{2}\delta^{\top}(t)\delta(t)=\frac{1}{2}y^{\top}(t)(K_n\otimes I_p)y(t).
\end{align}
Then its derivative along the trajectories of the multi-agent system (\ref{singley}) with event-triggered controller (\ref{uysingle}) is
\begin{align}
&\dot{V}(t)=y^{\top}(t)(K_n\otimes I_p)\dot{y}(t)\nonumber\\
=&y^{\top}(t)(K_n\otimes I_p)[-(DWD^{\top}\otimes I_p)y(t)+e(t)]\nonumber\\
=&-y^{\top}(t)(DWD^{\top}\otimes I_p)y(t)+\delta^{\top}(t)e(t)\nonumber\\
\le&-\frac{2}{\Delta_0}y^{\top}(t)(DD^{\top}\otimes I_p)y(t)+\beta_0\delta^{\top}(t)\delta(t)+\frac{1}{4\beta_0}\|e(t)\|^2\nonumber\\
\le&-2\beta_0y^{\top}(t)(K_n\otimes I_p)y(t)+\beta_0\delta^{\top}(t)\delta(t)+\frac{1}{4\beta_0}\|e(t)\|^2\nonumber\\
\le&-2\beta_0V(t)+\frac{n\alpha^2}{4\beta_0}e^{-2\beta t},
\end{align}
where the first inequality holds since $W$ is diagonal and its diagonal elements are all not smaller than $\frac{2}{\Delta_0}$ and the second inequality holds from Lemma \ref{lemma2}.
Hence
\begin{align}
V(t)\le& V(0)e^{-2\beta_0t}+\frac{n\alpha^2}{8\beta_0(\beta_0-\beta)}[e^{-2\beta t}-1]\label{betaV}\\
<&k_Ve^{-2\beta t},
\end{align}
where
\begin{align}\label{kV}
k_V=V(0)+\frac{n\alpha^2}{8\beta_0(\beta_0-\beta)}.
\end{align}
Thus
\begin{align}
\|y_i(t)-y_j(t)\|^2&\le2\|y_i(t)-\bar{y}(t)\|^2+2\|\bar{y}(t)-y_j(t)\|^2\nonumber\\
&\le4V(t)<4k_Ve^{-2\beta t}.\label{lengthysV}
\end{align}
Hence
\begin{align}
\lim_{t\to\infty}x_i(t)-x_j(t)=\lim_{t\to\infty}y_i(t)-\tau_i-(y_j(t)-\tau_j)=d_{ij},
\end{align}
exponentially.

{\bf (iii)} For $(i,j)\in\mathcal E(\mathcal G)$, define
\begin{align}
f_{ij}(l)=&\frac{2\Delta-2\|d_{ij}\|-l}{(\Delta-\|d_{ij}\|-l)^2}, l\in[0,\Delta-\|d_{ij}\|),\label{fij}\\
g_{ij}(l)=&\frac{2(\Delta-\|d_{ij}\|)^2}{(\Delta-\|d_{ij}\|-l)^3}, l\in[0,\Delta-\|d_{ij}\|),\label{gij}\\
h_{ij}(l)=&\frac{3\Delta-3\|d_{ij}\|-l}{(\Delta-\|d_{ij}\|-l)^3}, l\in[0,\Delta-\|d_{ij}\|).\label{hij}
\end{align}
We can easily check that both $f_{ij}(l)$, $g_{ij}(l)$ and $h_{ij}(l)$ are increasing functions on $[0,\Delta-\|d_{ij}\|)$. Then from (\ref{lengthys}), we have
\begin{align}\label{wij}
0<\omega_{ij}(t)<f_{ij}(k_{ij}), \forall t\ge0.
\end{align}
Then
\begin{align}
\|\dot{y}_i(t)\|=&\|e_i(t)-\sum_{j\in N_i}\omega_{ij}(t)(y_i(t)-y_j(t))\|\nonumber\\
\le&\|e_{i}(t)\|+\sum_{j\in N_i}\omega_{ij}(t)\|(y_i(t)-y_j(t))\|\nonumber\\
<&\alpha e^{-\beta t}+\sum_{j\in N_i}2f_{ij}(k_{ij})\sqrt{k_V}e^{-\beta t}.\label{dotyi}
\end{align}
From
\begin{align}
&\dot{e}_i(t)\nonumber\\
=&\sum_{j\in N_i}[\dot{\omega}_{ij}(t)(y_i(t)-y_j(t))+\omega_{ij}(t)(\dot{y}_i(t)-\dot{y}_j(t))]\nonumber\\
=&\sum_{j\in N_i}\Big\{h_{ij}(\|y_i(t)-y_j(t)\|)
\frac{(y_i(t)-y_j(t))^\top}{\|y_i(t)-y_j(t)\|}(\dot{y}_i(t)-\dot{y}_j(t))\nonumber\\
&\times(y_i(t)-y_j(t))+\omega_{ij}(t)(\dot{y}_i(t)-\dot{y}_j(t))\Big\},\label{dotei}
\end{align}
we have
\begin{align}
&\frac{d\|e_i(t)\|}{dt}\le\|\dot{e}_i(t)\|\nonumber\\
\le&\sum_{j\in N_i}g_{ij}(\|y_i(t)-y_j(t)\|)
(\|\dot{y}_i(t)\|+\|\dot{y}_j(t)\|)\nonumber\\
\le&\sum_{j\in N_i}g_{ij}(k_{ij})[\|\dot{y}_i(t)\|+\|\dot{y}_j(t)\|]<c_ie^{-\beta t},\label{dotnormei}
\end{align}
where
\begin{align}
c_i=&\sum_{j\in N_i}g_{ij}(k_{ij})\Big[2\alpha+\sum_{l\in N_i}2f_{il}(k_{il})\sqrt{k_V}\nonumber\\
&+\sum_{l\in N_j}2f_{jl}(k_{jl})\sqrt{k_V}\Big].
\end{align}
Thus, a sufficient condition to guarantee
\begin{align*}
\alpha e^{-\beta t}\ge\|e_i(t)\|=\int_{t_k^i}^t\frac{d\|e_i(s)\|}{ds}ds, \forall t\in[t_k^i,t_{k+1}^i),
\end{align*}
is
\begin{align*}
&\alpha e^{-\beta t}\ge\int_{t_k^i}^tc_ie^{-\beta s}ds=\frac{c_i}{\beta}[e^{-\beta t_k^i}-e^{-\beta t}]\\
\Leftrightarrow&(c_i+\alpha\beta)e^{-\beta t}\ge c_ie^{-\beta t_k^i}
\Leftrightarrow(c_i+\alpha\beta)e^{-\beta (t-t_k^i)}\ge c_i\\
\Leftarrow&(c_i+\alpha\beta)[1-\beta (t-t_k^i)]\ge c_i
\Leftrightarrow t-t^i_k\le\xi_i,
\end{align*}
where
\begin{align}\label{xi}
\xi_i=\frac{\alpha}{c_i+\alpha\beta}>0.
\end{align}
In other words, for all $t\in[t^i_k,t^i_k+\xi_i]$, $\|e_i(t)\|<\alpha e^{-\beta t}$ holds. Hence $t^i_{k+1}> t^i_k+\xi_i$.
\end{pf}

\begin{remark}\label{remarkbeta}
In order to determine $\beta$, $\beta_0$ should be known first. However $\beta_0$ is a global parameter since it relates to $\rho_2(DD^{\top})$ and $\Delta_0$. We can lower bound $\beta_0$ by $\frac{4}{n(n-1)\Delta}$ since $\Delta_0<\Delta$ and $\rho_2(DD^{\top})\ge\frac{4}{n(n-1)}$, see \cite{mohar1991eigenvalues}.
\end{remark}

Apparently, in order to check the inequality in the triggering condition (\ref{triggersingle}), continuous sensing of the relative positions between neighbors is needed, which is a drawback. In the following we will give an algorithm to avoid this. In other words, the following algorithm is an implementation of Theorem \ref{thm1} but it only requires agents to sense and to broadcast, at the triggering times. We name this algorithm as {\it Event-Triggered  Algorithm-1}, which is illustrated as follows.

For any agent $i$, at any time $s\ge0$, it knows its last triggering time $t^i_{k_i(s)}$ and its control input $u_i(s)=u_i(t^i_{k_i(s)})$ which is a constant until it determines its next triggering time. If agent $i$ also knows the relative position $x_i(s)-x_j(s)$ and $u_j(s)=u_j(t^j_{k_j(s)})$ which is a constant until agent $j$ determines its next triggering time, for $j\in N_i$. Then agent $i$ can predict
\begin{align}
x_i(t)-x_j(t)=&x_i(s)-x_j(s)\nonumber\\
&+(t-s)[u_i(t^i_{k_i(s)})-u_j(t^j_{k_j(s)})],\label{statexixj}
\end{align}
until $t\le\min\{t^i_{k_i(s)+1},t^j_{k_j(s)+1}\}$.
This means continuous sensing or broadcasting are not needed any more. Here, we name $\{t^i_{k_i(t)},u_i(t^i_{k_i(t)}),x_i(t^i_{k_i(t)})-x_j(t^i_{k_i(t)})\}$ as the {\it triggering information} that agent $i$ sends to agent $j$ for $j\in N_i$.
In conclusion, we propose the following algorithm.

\noindent {\it Event-Triggered  Algorithm-1 (ETA-1)}:
\begin{algorithmic}[1]
\State Choose $\alpha>0$ and $0<\beta<\beta_0$;
\State Initialize $t^{i}_1=0$ and $k=1$;
\State At time $s=t^{i}_k$, agent $i$ uses (\ref{statexixj}) to predict the relative position $x_i(t)-x_j(t),\forall j\in N_i$;
\State Agent $i$ substitutes these relative positions into $e_i(t)$ and finds out $\tau^i_{k+1}$ which is the smallest solution of equation $\|e_i(t)\|=\alpha e^{-\beta t}$;
\If {there exists $j\in N_i$ such that agent $j$ broadcasts its triggering information at $t_0\in(s,\tau^i_{k+1})$\footnote{This kind of situation can only occur at most finite times during $(s,\tau^i_{k+1})$ since $|N_i|$ is finite and  inter-event times of each agent are lower bounded by a positive constant.}}
 \State agent $i$ updates $s=t_0$ and goes back to Step 3;
\Else
\State agent $i$ determines $t^i_{k+1}=\tau^i_{k+1}$, updates its control input $u_i(t^i_{k+1})$ by sensing the relative positions between itself and its neighbors, broadcasts its triggering information to its neighbors, resets $k=k+1$, and goes back to Step 3;
\EndIf
\end{algorithmic}

When applying above algorithm, although continuous sensing and broadcasting are avoided, each agent still needs to continuously listen for incoming information from its neighbors since the triggering times are not known in advance. Our future work will focus on proposing self-triggered algorithms to avoid this.

\section{Formation Control with Connectivity Preservation: Double Integrators}\label{secdouble}
In this section, we extend the results in above section to the case when the dynamics of agents can be modeled as double integrators given by
\begin{align}\label{double}
\begin{cases}
\dot{x}_i(t)=q_i(t),\\
\dot{q}_i(t)=u_i(t),~i\in\mathcal V,
\end{cases}
\end{align}
where $x_i(t)\in\mathbb R^p$ still denotes the position of agent $i$ at time $t$, $q_i(t)\in\mathbb R^p$ denotes the speed and $u_i(t)\in\mathbb R^p$ is the control input.
We can also rewrite (\ref{double}) as
\begin{align}\label{doubley}
\begin{cases}
\dot{y}_i(t)=q_i(t),\\
\dot{q}_i(t)=u_i(t),~i\in\mathcal V.
\end{cases}
\end{align}
Denote
\begin{align*}
C=\left[\begin{array}{rr}0&1\\
0&0
\end{array}\right],
B=\left[\begin{array}{r}0\\
1
\end{array}\right],
z_i(t)=\left[\begin{array}{r}y_i(t)\\
q_i(t)
\end{array}\right],
\end{align*}
then we can rewrite (\ref{doubley}) as
\begin{align*}
\dot{z}_i(t)=(C\otimes I_p)z_i(t)+(B\otimes I_p)u_i(t),
\end{align*}
One can easily check that $(C,B)$ is controllable and $(I_2,C)$ is observable. Hence, from \cite{Kucera1972}, we know that there exist positive constants $k_0,k_1$ and $k_2$ such that
\begin{align}\label{riccatti}
P>0,~\frac{1}{2}(PC+C^{\top}P)-\beta_1 PBB^{\top}P+2I_2\le0,
\end{align}
with
$
P=\left[\begin{array}{rr}k_0&k_1\\
k_1&k_2
\end{array}\right]
$ and $0<\beta_1\le \beta_0$.

Similar to the event-triggered controller (\ref{uxsingle}), we use the following event-triggered controller
\begin{align}
&u_i(t)\nonumber\\
=&-k_1\sum_{j\in N_i}\omega_{ij}(t^i_{k_i(t)})(y_i(t^i_{k_i(t)})-y_j(t^i_{k_i(t)}))\nonumber\\
&-k_2\sum_{j\in N_i}\omega_{ij}(t^i_{k_i(t)})(q_i(t^i_{k_i(t)})-q_j(t^i_{k_i(t)}))-k_3q_i(t)\label{uydouble}\\
=&-k_1\sum_{j\in N_i}\omega_{ij}(t^i_{k_i(t)})(x_i(t^i_{k_i(t)})-x_j(t^i_{k_i(t)})-d_{ij})\nonumber\\
&-k_2\sum_{j\in N_i}\omega_{ij}(t^i_{k_i(t)})(q_i(t^i_{k_i(t)})-q_j(t^i_{k_i(t)}))-k_3q_i(t),\label{uxdouble}
\end{align}
where $k_3$ is a constant which will be determined later. Here we should highlight that this controller needs continuous feedback because of the item $k_3q_i(t)$.

Similar to Theorem \ref{thm1}, we have the following results.
\begin{theorem}\label{thm2}
Given a graph $\mathcal G$ which is undirected and connected and a formation associated with $\mathcal G$ which satisfies Assumption \ref{assum1}. Given $0<\beta_1\le\beta_0$, determine $P$ by (\ref{riccatti}).
Consider the multi-agent system (\ref{double}) with event-triggered controller (\ref{uxdouble}) associated with $\mathcal G$.
Assume the initial position condition is also satisfied (\ref{initial})
for all $(i,j)\in\mathcal E(\mathcal G)$. Given $0<k_3<\frac{4}{k_2+\sqrt{k_1^2+k_2^2}}$, and given $\alpha_d>0$, and $0<\beta_d<(2-k_4)\rho_2(P)$ with $k_4=k_3\frac{k_2+\sqrt{k_1^2+k_2^2}}{2}<2$, and given the first triggering time $t^i_1=0$, agent $i$ determines the triggering times $\{t^i_k\}_{k=2}^{\infty}$ by
\begin{align}\label{triggerdouble}
t^i_{k+1}=\max_{r\ge t^i_k}\{r:\|E_i(t)\|\le \alpha_d e^{-\beta_d t},\forall t\in[t^i_k,r]\}
\end{align}
where
\begin{align*}
E_{i}(t)=&k_1e_i(t)+k_2e_{qi}(t),\\
e_{qi}(t)=&\sum_{j\in N_i}\omega_{ij}(t)(q_i(t)-q_j(t))\\
&-\sum_{j\in N_i}\omega_{ij}(t^i_{k_i(t)})(q_i(t^i_{k_i(t)})-q_j(t^i_{k_i(t)})).
\end{align*}
Then the multi-agent system (\ref{double}) with event-triggered controller (\ref{uxdouble}) converges to the formation  exponentially with connectivity preservation, and there is no Zeno behavior.
\end{theorem}
\begin{pf}
This theorem holds if we can prove that
\begin{description}
  \item[(i)] $\|x_i(t)-x_j(t)\|\le\Delta, \forall (i,j)\in\mathcal E(\mathcal G), \forall t\ge0$;
  \item[(ii)] $\lim_{t\to\infty}x_i(t)-x_j(t)=d_{ij}, \forall(i,j)\in\mathcal E(\mathcal G)$,  exponentially;
  \item[(iii)] there exists $\xi_i^d>0$, such that $t^i_{k+1}-t^i_k\ge\xi^d_i$, $\forall i\in\mathcal V$ and $\forall k=1,2,\dots$;
  \item[(iv)] $\lim_{t\to\infty}q_i(t)=0$, $\forall i\in\mathcal V$.
\end{description}

{\bf (i)} We define the total tension energy of $\mathcal G$ as
\begin{align}
\nu_d(\Delta,y)=k_1\nu(\Delta,y)+\frac{1}{2}\sum_{i=1}^{n}\|q_i(t)\|^2.
\end{align}
Then time derivative of $\nu_d(\Delta,y(t))$ along  the trajectories of the multi-agent system (\ref{doubley}) with event-triggered controller (\ref{uydouble}) is
\begin{align}
&\dot{\nu}_d(\Delta,y(t))\nonumber\\
=&k_1\sum_{i=1}^{n}\sum_{j\in N_i}\Big[\frac{\partial\nu_{ij}(\Delta,y)}{\partial y_i}\Big]^{\top}\Big|_{y=y(t)}\dot{y}_i(t)+\sum_{i=1}^{n}q^{\top}_i(t)\dot{q}_i(t)\nonumber\\
=&\sum_{i=1}^{n}q^{\top}_i(t)\Big\{k_1\sum_{j\in N_i}[\omega_{ij}(t)(y_i(t)-y_j(t))]+u_i(t)\Big\}\nonumber\\
=&\sum_{i=1}^{n}q^{\top}_i(t)\Big\{E_i(t)-k_2\sum_{j\in N_i}\omega_{ij}(t)(q_i(t)-q_j(t))-k_3q_i(t)\Big\}\nonumber\\
\le&-\sum_{i=1}^{n}(k_3-b)q^{\top}_i(t)q_i(t)+\frac{1}{4b}\sum_{i=1}^{n}\|E_i(t)\|^2\nonumber\\
&-k_2q^{\top}(t)L_{\omega}q(t),
\end{align}
for all $0<b<k_3$. From (\ref{triggerdouble}), we know that
\begin{align}
\|E_i(t)\|\le\alpha_d e^{-\beta_d t}, \forall t\ge0.
\end{align}
Hence
\begin{align*}
\dot{\nu}_d(\Delta,y(t))\le\frac{n\alpha^2_d}{4b}e^{-2\beta_d t}, \forall t\ge0.
\end{align*}
Thus
\begin{align}
\nu_d(\Delta,y(t))\le&\nu_d(\Delta,y(0))+\frac{n\alpha^2_d}{8b\beta_d}[1-e^{-2\beta_d t}]\nonumber\\
\le&k^d_{\nu}, \forall t\ge0,
\end{align}
where
\begin{align}
&k^d_{\nu}=\nu_d(\Delta,y(0))+\frac{n\alpha_d^2}{8b\beta_d}\nonumber\\
=&\frac{k_1}{2}\sum_{i=1}^{n}\sum_{j\in N_i}\frac{\|x_i(0)-x_j(0)-d_{ij}\|^2}{\Delta-\|d_{ij}\|-\|x_i(0)-x_j(0)-d_{ij}\|}\nonumber\\
&+\frac{1}{2}\sum_{i=1}^{n}\|q_i(0)\|^2+\frac{n\alpha_d^2}{8b\beta_d}.
\end{align}
Then, for any $(i,j)\in\mathcal E(\mathcal G)$ and $t\ge0$, we have
\begin{align*}
\nu_{ij}(\Delta,y(t))&=\frac{\|y_i(t)-y_j(t)\|^2}{\Delta-\|d_{ij}\|-\|y_i(t)-y_j(t)\|}\\
&\le\frac{2}{k_1}\nu_d(\Delta,y(t))\le \frac{2}{k_1}k^d_{\nu}.
\end{align*}
Hence
\begin{align}\label{lengthyd}
\|y_i(t)-y_j(t)\|\le k^d_{ij},
\end{align}
where
\begin{align}\label{kijd}
k^d_{ij}=-\frac{k^d_\nu}{k_1}+\sqrt{\Big(\frac{k^d_\nu}{k_1}\Big)^2+2\frac{k^d_\nu}{k_1}(\Delta-\|d_{ij}\|)}<\Delta-\|d_{ij}\|.
\end{align}
Then, we have
\begin{align}
&\|x_i(t)-x_j(t)\|=\|x_i(t)-\tau_i-(x_j(t)-\tau_j)+d_{ij}\|\nonumber\\
=&\|y_i(t)-y_j(t)+d_{ij}\|\le\|y_i(t)-y_j(t)\|+\|d_{ij}\|\nonumber\\
\le&k^d_{ij}+\|d_{ij}\|<\Delta,
\end{align}
and thus connectivity maintenance is guaranteed.

{\bf (ii)}  Noting 
$
B^{\top}P=\left[\begin{array}{rr}k_1&k_2
\end{array}\right]
$,
we can rewrite the control input (\ref{uydouble}) as
\begin{align*}
u_i(t)=&-(B^{\top}P\otimes I_p)\sum_{j\in N_i}\omega_{ij}(t)(z_i(t)-z_j(t))+E_i(t)\\
&-k_3(B^{\top}\otimes I_p)z_i(t).
\end{align*}
Let
 $z(t)=(z_1^{\top}(t),\dots,z_n^{\top}(t))^{\top}$,
 and $\bar{z}(t)=\frac{1}{n}\sum_{i=1}^{n}z_i(t)$.
We consider the following candidate Lyapunov function
\begin{align}
V_d(t)=&\frac{1}{2}[z(t)-{\bf1}_n\bar{z}(t)]^{\top}(I_n\otimes P\otimes I_p)[z(t)-{\bf1}_n\bar{z}(t)]\nonumber\\
=&\frac{1}{2}z^{\top}(t)(K_n\otimes P\otimes I_p)z(t).
\end{align}
Then its derivative is
\begin{align}
&\dot{V}_d(t)=z^{\top}(t)(K_n\otimes P\otimes I_p)\dot{z}(t)\nonumber\\
=&z^{\top}(t)(K_n\otimes P\otimes I_p)\Big\{(I_n\otimes C\otimes I_p)z(t)\nonumber\\
&+(I_n\otimes B\otimes I_p)u(t)\Big\}\nonumber\\
=&z^{\top}(t)(K_n\otimes P\otimes I_p)\Big\{(I_n\otimes C\otimes I_p)z(t)\nonumber\\
&+(I_n\otimes B\otimes I_p)[-(L_{\omega}\otimes B^{\top}P\otimes I_p)z(t)+E(t)\nonumber\\
&-k_3(I_n\otimes B^{\top}\otimes I_p)z(t)]\Big\}\nonumber\\
=&z^{\top}(t)(K_n\otimes \frac{PC+CP^{\top}}{2}\otimes I_p)z(t)\nonumber\\
&-z^{\top}(t)(L_{\omega}\otimes PBB^{\top}P\otimes I_p)z(t)\nonumber\\
&-k_3z^{\top}(t)(K_n\otimes PBB^{\top}\otimes I_p)z(t)\nonumber\\
&+z^{\top}(t)(K_n\otimes PB\otimes I_p)E(T),
\end{align}
where $E(t)=[E_1^{\top}(t),\dots,E_n^{\top}(t)]$.
Noting $PBB^{\top}P\ge0$ and $L_{\omega}\ge2\beta_0K_n\ge2\beta_1 K_n$ from Lemma \ref{lemma2}, we have
\begin{align*}
&-z^{\top}(t)(L_{\omega}\otimes PBB^{\top}P\otimes I_p)z(t)\\
\le&-2\beta_1 z^{\top}(t)(K_n\otimes PBB^{\top}P\otimes I_p)z(t).
\end{align*}
Noting
\begin{align*}
\frac{PBB^{\top}+BB^{\top}P}{2}=\left[\begin{array}{cc}0&\frac{k_1}{2}\\
\frac{k_1}{2}&k_2
\end{array}\right],
\end{align*}
one can easily check that $\rho(\frac{PBB^{\top}+BB^{\top}P}{2})=\frac{k_2+\sqrt{k_1^2+k_2^2}}{2}$.
Noting $k_4=k_3\frac{k_2+\sqrt{k_1^2+k_2^2}}{2}$, we have
\begin{align*}
&-k_3z^{\top}(t)(K_n\otimes PBB^{\top}\otimes I_p)z(t)\\
\le&k_4z^{\top}(t)(K_n\otimes I_2\otimes I_p)z(t).
\end{align*}
Then from above two inequalities and the following inequality
\begin{align*}
&z^{\top}(t)(K_n\otimes PB\otimes I_p)E(T)\\
\le&\beta_1 z^{\top}(t)(K_n\otimes PBB^{\top}P\otimes I_p)z(t)+\frac{1}{4\beta_1}\|E(t)\|^2,
\end{align*}
we get
\begin{align}
&\dot{V}_d(t)\nonumber\\
\le& z^{\top}(t)\Big(K_n\otimes \Big[\frac{PC+CP^{\top}}{2}-\beta_1 PBB^{\top}P\Big]\otimes I_p\Big)\nonumber\\
&+k_4z^{\top}(t)(K_n\otimes I_2\otimes I_p)z(t)+\frac{1}{4\beta_1}\|E(t)\|^2\nonumber\\
\le&-(2-k_4)z^{\top}(t)(K_n\otimes I_2\otimes I_p)z(t)+\frac{1}{4\beta_1}\|E(t)\|^2\nonumber\\
\le&-2(2-k_4)\rho_2(P)V_d(t)+\frac{n\alpha_d^2}{4\beta_1}e^{-2\beta_d t},
\end{align}
where the second inequality holds from (\ref{riccatti}).
Hence
\begin{align}
V_d(t)
\le& V_d(0)e^{-2(2-k_4)\rho_2(P)t}\nonumber\\
&+\frac{n\alpha_d^2}{8\beta_1[(2-k_4)\rho_2(P)-\beta_d]}[e^{-2\beta_dt}-1]\nonumber\\
<&k^d_Ve^{-2\beta_d t},
\end{align}
where
\begin{align}\label{kVd}
k^d_V=V_d(0)+\frac{n\alpha_d^2}{8\beta_1[(2-k_4)\rho_2(P)-\beta_d]}.
\end{align}
Thus
\begin{align}
&\|y_i(t)-y_j(t)\|^2+\|q_i(t)-q_j(t)\|^2=\|z_i(t)-z_j(t)\|^2\nonumber\\
&\le2\|z_i(t)-\bar{z}(t)\|^2+2\|\bar{z}(t)-z_j(t)\|^2\nonumber\\
&\le4V_d(t)<4k^d_Ve^{-2\beta_d t}.\label{yiyj}
\end{align}
Hence
\begin{align}\label{xfinald}
\lim_{t\to\infty}x_i(t)-x_j(t)=\lim_{t\to\infty}y_i(t)-\tau_i-(y_j(t)-\tau_j)=d_{ij},
\end{align}
and
\begin{align}\label{qfinald}
\lim_{t\to\infty}q_i(t)-q_j(t)=0,
\end{align}
exponentially.

{\bf (iii)}
Similar to (\ref{dotyi}), we have
\begin{align}
&\|\dot{q}_i(t)-\dot{q}_j(t)\|\nonumber\\
=&\|E_i(t)-E_{j}(t)-k_1\sum_{l\in N_i}\omega_{il}(t)(y_i(t)-y_l(t))\nonumber\\
&-k_2\sum_{l\in N_i}\omega_{il}(t)(q_i(t)-q_l(t))\nonumber\\
&+k_1\sum_{l\in N_j}\omega_{jl}(t)(y_j(t)-y_l(t))\|\nonumber\\
&+k_2\sum_{l\in N_j}\omega_{jl}(t)(q_j(t)-q_l(t))-k_3(q_i(t)-q_j(t))\|\label{dotqie}\\
<&-k_3(q_i(t)-q_j(t))\|c^q_{ij}e^{-\beta_d t},\label{dotqi}
\end{align}
where
\begin{align}
c^q_{ij}=&2\alpha_d+2\Big\{(k_1+k_2)\Big[\sum_{l\in N_i}f_{il}(k^d_{il})+\sum_{l\in N_j}f_{jl}(k^d_{jl})\Big]\nonumber\\
&+k_3\Big\}\sqrt{k^d_V}.\label{cijq}
\end{align}
Similar to (\ref{dotei}), we have
\begin{align}
&\dot{e}_i(t)\nonumber\\
=&\sum_{j\in N_i}[\dot{\omega}_{ij}(t)(y_i(t)-y_j(t))+\omega_{ij}(t)(\dot{y}_i(t)-\dot{y}_j(t))]\nonumber\\
=&\sum_{j\in N_i}\Big\{h_{ij}(\|y_i(t)-y_j(t)\|)
\frac{(y_i(t)-y_j(t))^\top}{\|y_i(t)-y_j(t)\|}(q_i(t)-q_j(t))\nonumber\\
&\times(y_i(t)-y_j(t))+\omega_{ij}(t)(q_i(t)-q_j(t))\Big\},\label{doteidouble}
\end{align}
and
\begin{align}
&\dot{e}_{qi}(t)\nonumber\\
=&\sum_{j\in N_i}[\dot{\omega}_{ij}(t)(q_i(t)-q_j(t))+\omega_{ij}(t)(\dot{q}_i(t)-\dot{q}_j(t))]\nonumber\\
=&\sum_{j\in N_i}\Big\{h_{ij}(\|y_i(t)-y_j(t)\|)
\frac{(y_i(t)-y_j(t))^\top}{\|y_i(t)-y_j(t)\|}(q_i(t)-q_j(t))\nonumber\\
&\times(q_i(t)-q_j(t))+\omega_{ij}(t)(\dot{q}_i(t)-\dot{q}_j(t))\Big\}.\label{doteqi}
\end{align}
Similar to (\ref{dotnormei}), we have
\begin{align}
&\frac{d\|k_1e_i(t)+k_2e_{qi}(t)\|}{dt}\le k_1\|\dot{e}_i(t)\|+k_2\|\dot{e}_{qi}(t)\|\nonumber\\
\le&\sum_{j\in N_i}\Big\{k_1g_{ij}(\|y_i(t)-y_j(t)\|)
\|q_i(t)-q_j(t)\|\nonumber\\
&+k_2h_{ij}(\|y_i(t)-y_j(t)\|)\|q_i(t)-q_j(t)\|^2\nonumber\\
&+k_2\omega_{ij}(t)(\|\dot{q}_i(t)-\dot{q}_j(t)\|)\Big\}\label{doteq}\\
\le&\sum_{j\in N_i}k_1g_{ij}(k^d_{ij})\|q_i(t)-q_j(t)\|+k_2h_{ij}(k^d_{ij})\|q_i(t)-q_j(t)\|^2\nonumber\\
&+k_2f_{ij}(k^d_{ij})(\|\dot{q}_i(t)-\dot{q}_j(t)\|)
<c^d_ie^{-\beta_d t},
\end{align}
where
\begin{align}
c^d_i=&\sum_{j\in N_i}\Big\{2k_1g_{ij}(k^d_{ij})\sqrt{k^d_V}+4k_2h_{ij}(k^d_{ij})k^d_V\nonumber\\
&+k_2f_{ij}(k^d_{ij})c^q_{ij}\Big\}.
\end{align}
Similar to the way to find $\xi_i$, we can find
\begin{align}\label{xid}
\xi_i^d=\frac{\alpha_d}{c^d_i+\alpha_d\beta_d}>0,
\end{align}
such that $t^i_{k+1}> t^i_k+\xi^d_i$.

{\bf (iv)} From the above result we have $\lim_{t\to\infty}t^i_k=\infty$. Then from (\ref{xfinald}) and (\ref{qfinald}), we know that $\lim_{t\to\infty}u_i(t)=-k_3\lim_{t\to\infty}q_i(t)$. Thus $\lim_{t\to\infty}q_i(t)=0$.
\end{pf}

Similar to the analysis after Theorem \ref{thm1}, in order to check the inequality in the triggering condition (\ref{triggerdouble}), continuous sensing of the relative positions and speeds between neighbors is needed. In the following we will give an algorithm to implement Theorem \ref{thm2} and at the same time to avoid continuous sensing by using the same idea as ETA-1.

Denote
\begin{align*}
u^d_i(t)=&-k_1\sum_{j\in N_i}\omega_{ij}(t^i_{k_i(t)})(x_i(t^i_{k_i(t)})-x_j(t^i_{k_i(t)})-d_{ij})\\
&-k_2\sum_{j\in N_i}\omega_{ij}(t^i_{k_i(t)})(q_i(t^i_{k_i(t)})-q_j(t^i_{k_i(t)})),
\end{align*}
which is piecewise constant.

For any agent $i$, at any time $s\ge0$, it knows its last triggering time $t^i_{k_i(s)}$, its current speed $q_i(s)$, and  $u^d_i(t^i_{k_i(s)})$ which is a constant until it determines its next triggering time. If agent $i$ also knows the relative position $x_i(s)-x_j(s)$, relative speed $q_i(s)-q_j(s)$ and $u^d_j(s)=u^d_j(t^j_{k_j(s)})$ which is a constant until agent $j$ determines its next triggering time, for $j\in N_i$.  Then agent $i$ can predict
\begin{align}
&x_i(t)-x_j(t)=\frac{1}{k_3}[1-e^{k_3(s-t)}][q_i(s)-q_j(s)]\nonumber\\
&+[k_3(t-s)-1+e^{k_3(s-t)}]\frac{u^d_i(t^i_{k_i(s)})-u^d_j(t^j_{k_j(s)})}{(k_3)^2},\label{statexixjd}\\
&q_i(t)-q_j(t)=e^{k_3(s-t)}[q_i(s)-q_j(s)]\nonumber\\
&+[1-e^{k_3(s-t)}]\frac{u^d_i(t^i_{k_i(s)})-u^d_j(t^j_{k_j(s)})}{k_3},\label{stateqiqj}
\end{align}
until $t\le\min\{t^i_{k_i(s)+1},t^j_{k_j(s)+1}\}$.
This means continuous sensing or broadcasting are not needed any more.
We name $\{t^i_{k_i(t)},x_i(t^i_{k_i(t)})-x_j(t^i_{k_i(t)}),q_i(t^i_{k_i(t)})-q_j(t^i_{k_i(t)}),u^d_i(t^i_{k_i(t)})\}$ as the {\it triggering information} that agent $i$ sends to agent $j$ for $j\in N_i$.
In conclusion, we propose the following algorithm.

\noindent {\it Event-Triggered Algorithm-2 (ETA-2)}:
\begin{algorithmic}[1]
\State Choose $0<\beta_1\le\beta_0$ and determine $P$ by (\ref{riccatti});
\State Choose $0<k_3<\frac{4}{k_2+\sqrt{k_1^2+k_2^2}}$, $\alpha_d>0$ and $0<\beta_d<(2-k_4)\rho_2(P)$;
\State Initialize $t^{i}_1=0$ and $k=1$;
\State At time $s=t^{i}_k$, agent $i$ uses (\ref{statexixjd}) and (\ref{stateqiqj})  to predict the relative position $x_i(t)-x_j(t)$ and the relative speed $q_i(t)-q_j(t),\forall j\in N_i$;
\State Agent $i$ substitutes these into $E_i(t)$ and finds out $\tau^i_{k+1}$ which is the smallest solution of equation $\|E_i(t)\|=\alpha_d e^{-\beta_d t}$;
\If {there exists $j\in N_i$ such that agent $j$ broadcasts its triggering information at $t_0\in(s,\tau^i_{k+1})$}
 \State agent $i$ updates $s=t_0$ and goes back to Step 4;
\Else
\State agent $i$ determines $t^i_{k+1}=\tau^i_{k+1}$, updates  $u^d_i(t^i_{k+1})$ by sensing the relative positions and relative speeds between itself and its neighbors, broadcasts its triggering information to its neighbors, resets $k=k+1$, and goes back to Step 4;
\EndIf
\end{algorithmic}

As noted earlier, each agent still needs to continuously listen for incoming information from
its neighbors, which will be our future work.

\section{Simulations}\label{secsimulations}
In this section, two numerical examples are given to demonstrate the effectiveness of the presented results.

Consider a network of $n=3$ agents in $\mathbb{R}^2$ whose Laplacian matrix is given by
\begin{eqnarray*}
L=\left[\begin{array}{ccc}2&-1&-1\\
-1&2&-1\\
-1&-1&2
\end{array}\right].
\end{eqnarray*}
The three agents are trying to establish a right triangle formation with
\begin{align*}
d_{12}=\left(\begin{array}{c}0\\
-2
\end{array}\right),
d_{13}=\left(\begin{array}{c}-2\\
0
\end{array}\right),
d_{23}=\left(\begin{array}{c}-2\\
2
\end{array}\right).
\end{align*}
The communication radius is $\Delta=4$. We have $\beta_0=1.5$.

\begin{figure}[hbt]
\begin{subfigure}{.5\textwidth}
  \centering
  \includegraphics[width=.9\linewidth]{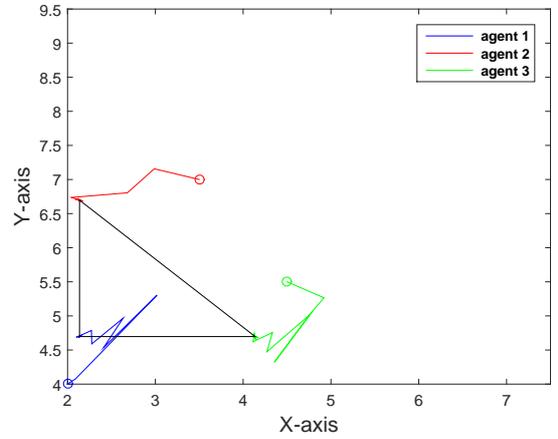}
  \caption{}
  \label{fig:1a}
\end{subfigure}%
\\
\begin{subfigure}{.5\textwidth}
  \centering
  \includegraphics[width=.9\linewidth]{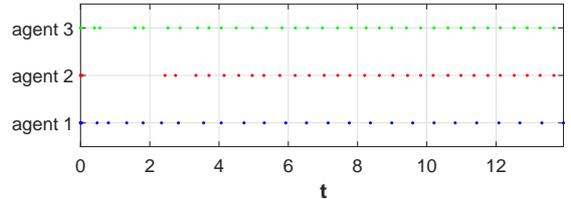}
  \caption{}
  \label{fig:1b}
\end{subfigure}
\caption{(a) The trajectories evolution of the multi-agent system (\ref{single}) with event-triggered controller (\ref{uxsingle}) when performing ETA-1. (b) The triggering times of each agent.}
\label{fig:1}
\end{figure}

Firstly, we consider the situation that the three agents are modeled as single integrators.
The initial positions of agents can be randomly selected as long as the initial condition (\ref{initial}) can be satisfied. Here, the initial position of agents are given by
\begin{align*}
x_1(0)=\left(\begin{array}{c}2\\
4
\end{array}\right),
x_2(0)=\left(\begin{array}{c}3.5\\
7
\end{array}\right),
x_3(0)=\left(\begin{array}{c}4.5\\
5.5
\end{array}\right).
\end{align*}
One can easily check that both Assumption \ref{assum1} and initial condition (\ref{initial}) hold. Choose $\alpha=10$ and $\beta=1$, by applying the ETA-1, we get the evolution of the formation shown in Fig. \ref{fig:1} (a), where ``circles'' denote the initial positions and ``triangle'' denotes the desired formation, and the triggering times of each agent shown in Fig. \ref{fig:1} (b), respectively.

\begin{figure}[hbt]
\begin{subfigure}{.5\textwidth}
  \centering
  \includegraphics[width=.9\linewidth]{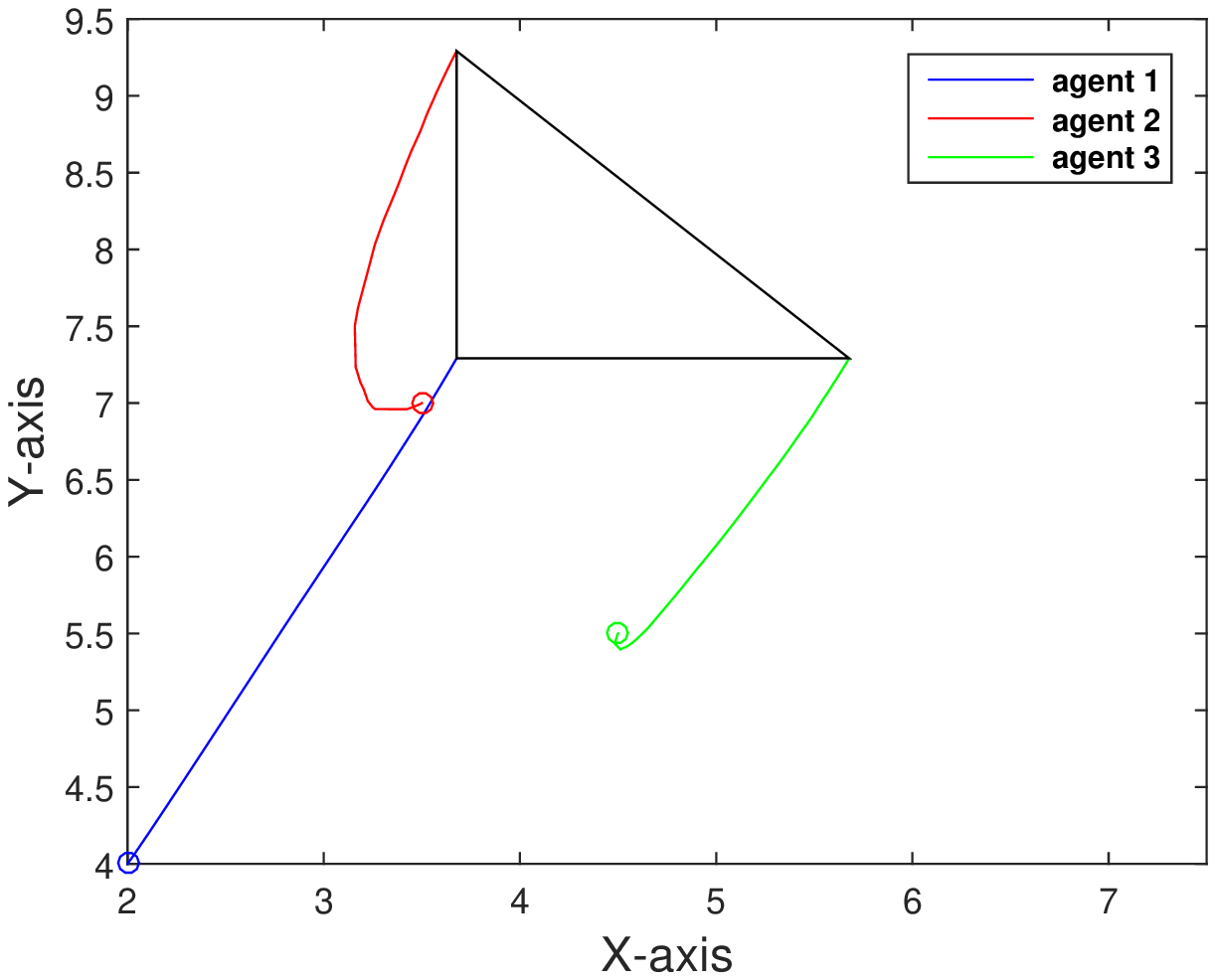}
  \caption{}
  \label{fig:2a}
\end{subfigure}%
\\
\begin{subfigure}{.5\textwidth}
  \centering
  \includegraphics[width=.9\linewidth]{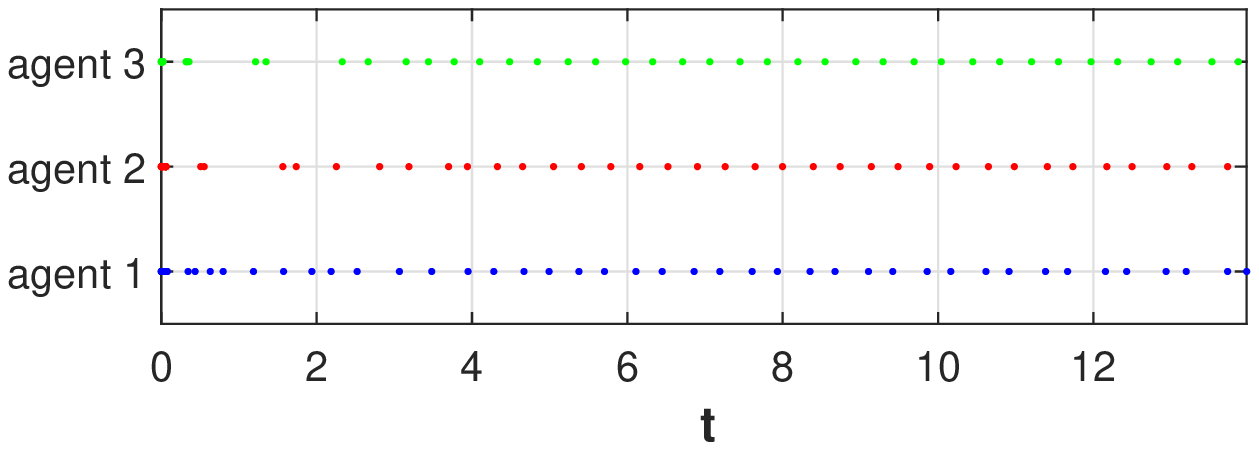}
  \caption{}
  \label{fig:2b}
\end{subfigure}
\caption{(a) The trajectories evolution of the multi-agent system (\ref{double}) with event-triggered controller (\ref{uxdouble}) when performing ETA-2. (b) The triggering times of each agent.}
\label{fig:2}
\end{figure}

Secondly, we consider the situation that the three agents are modeled as double integrators.
The initial positions of agents are given as before. The initial speeds of agents can be randomly selected and here we choose
\begin{align*}
q_1(0)=\left(\begin{array}{c}1\\
2
\end{array}\right),
q_2(0)=\left(\begin{array}{c}-1\\
-2
\end{array}\right),
q_3(0)=\left(\begin{array}{c}-1\\
-1
\end{array}\right).
\end{align*}
We have $P=\left[\begin{array}{cc}5.0237&1.1547\\1.1547&1.4502\end{array}\right]$, $k_1=1.1547$, $k_2=1.4502$, and $\rho_2(P)=1.1096$. Choose $k_3=\frac{2}{k_2+\sqrt{k_1^2+k_2^2}}=0.6053$, $\alpha_d=10$, and $\beta_d=1$, by applying the ETA-2, we get the evolution of the formation shown in Fig. \ref{fig:2} (a), where ``circles'' denote the initial positions and ``triangle'' denotes the desired formation, and the triggering times of each agent shown in Fig. \ref{fig:2} (b), respectively.  It can be seen that double integrators have more smooth trajectories compared with single integrators.

\section{Conclusions}\label{secconclusions}
In this paper, formation control for multi-agent systems with limited sensing and broadcasting radius is addressed. We first consider the situation that agents are modeled as single integrators. An event-triggered controller and a corresponding algorithm, to avoid continuous sensing and broadcasting, are proposed. Each agent only needs to sense and to broadcast at its triggering times, and to receive at its neighbors' triggering times. As a result, the desired formation can be established exponentially with connectivity preservation and exclusion of Zeno behavior. Then, these results are extended to double integrators. The drawback of our approach is that  each agent still needs to continuously
listen for incoming information from its neighbors.  Future research directions of this work include developing
self-triggered algorithm to avoid this.

\bibliography{references}             








\appendix

\end{document}